%% file: main.tex
\documentclass[aip,amsmath,amssymb,reprint]{revtex4-1}

\usepackage{graphicx}
\usepackage{dcolumn}
\usepackage{bm}
\usepackage[dvipsnames]{xcolor}
\usepackage{soul}
\usepackage{textcomp}

\newcommand{\review}[1]{\textcolor{black}{#1}}

\usepackage[utf8]{inputenc}
\usepackage[T1]{fontenc}
\usepackage{mathptmx}

\draft 

\begin{document}



\title{Operation of parallel SNSPDs at high detection rates} 

\author{Matthieu~Perrenoud}
\email[]{matthieu.perrenoud@unige.ch}
\affiliation{Group of Applied Physics, University of Geneva, CH-1211 Geneva, Switzerland}

\author{Misael~Caloz}
\affiliation{Group of Applied Physics, University of Geneva, CH-1211 Geneva, Switzerland}

\author{Emna~Amri}
\affiliation{Group of Applied Physics, University of Geneva, CH-1211 Geneva, Switzerland}
\affiliation{ID Quantique SA, CH-1227 Geneva, Switzerland}

\author{Claire~Autebert}
\affiliation{Group of Applied Physics, University of Geneva, CH-1211 Geneva, Switzerland}

\author{Christian~Sch{\"o}nenberger}
\affiliation{Department of Physics, University of Basel, CH-4056 Basel, Switzerland}

\author{Hugo~Zbinden}
\affiliation{Group of Applied Physics, University of Geneva, CH-1211 Geneva, Switzerland}

\author{Félix~Bussières}
\affiliation{Group of Applied Physics, University of Geneva, CH-1211 Geneva, Switzerland}
\affiliation{ID Quantique SA, CH-1227 Geneva, Switzerland}

\date{\today}


\begin{abstract}
Recent progress in the development of superconducting nanowire single-photon detectors (SNSPD) has delivered excellent performance, and their increased adoption has had a great impact on a range of applications. One of the key characteristic of SNSPDs is their detection rate, which is typically higher than other types of free-running single-photon detectors. The maximum achievable rate is limited by the detector recovery time after a detection, which itself is linked to the superconducting material properties and to the geometry of the meandered SNSPD. Arrays of detectors biased individually can be used to solve this issue, but this approach significantly increases both the thermal load in the cryostat and the need for time processing of the many signals, and this scales unfavorably with a large number of detectors. One potential scalable approach to increase the detection rate of individual detectors further is based on parallelizing smaller meander sections. 
In this way, a single detection temporarily disables only one subsection of the whole active area, thereby leaving the overall detection efficiency mostly unaffected.
In practice however, cross-talk between parallel nanowires typically leads to latching, which prevents high detection rates. Here we show how this problem can be avoided through a careful design of the whole SNSPD structure. \review{Using the same electronic readout as with conventional SNSPDs and a single coaxial line, we demonstrate detection rates over 200 MHz without any latching, and a fibre-coupled SDE as high as 77\%, and more than 50\% average SDE per photon at 50 MHz detection rate under continuous wave illumination.}

\end{abstract}


\maketitle 


Superconducting nanowire single-photon detectors\cite{Gol2001} (SNSPDs) are known for yielding overall excellent performance thanks to their high efficiency\cite{Marsili2013}, low dark count rate of less than one Hertz\cite{Shibata2015}, ultra-short dead time\cite{Vetter2016} as well as timing jitter in the range of a few to tens of picoseconds\cite{Caloz2018, Zadeh2017, Korzh2018, Caloz2019}. This makes them a key technology for application such as optical quantum information processing\cite{Hadfield2009}, deep-space optical communication\cite{Grein2015} and optical quantum computing\cite{Qiang2018}. In particular, their typical dead time in the range of tens of nanoseconds is of great interest in various applications requiring high detection rates with free-running detectors such as quantum key distribution \cite{Takesue2007, Boaron2018a}.

The most common SNSPDs based on a single meander yields detection rates of tens of MHz. Indeed, this rate is ultimately limited by the recovery time of the biasing current\cite{Kerman2006}, i.e.~when a photon is absorbed the nanowire becomes resistive and the biasing current rapidly leaves the nanowire (with a typical time of about 1~ns or less), this effectively leads to a zero efficiency right after the detection. After the current left the nanowire, it rapidly cools back to its superconducting state, and the current flows back inside the nanowire. At this point, the kinetic inductance of the superconducting meander forces the bias current to recover with a time constant $\tau = L_k/R$, where $R$ is the overall series impedance of the readout circuit and $L_k$ is the kinetic inductance of the nanowire. The timing constant $\tau$ typically is in the range of a few tens of nanoseconds for a single meander. A longer nanowire exhibits a larger kinetic inductance, which slows its current dynamics down and hence increases the time needed before the detector recovers its nominal efficiency. Consequently, at high detection rates where successive detections can happen in time intervals of the order of $\tau$, the system detection efficiency (SDE) drops. \review{Shorter recovery times, hence higher detection rates, can be obtained by reducing the kinetic inductance of the detector. One approach is cascade switching SNSPDs\cite{Ejrnaes2007}, also known as SNAPs, however their efficiency after detection still drops to zero as with conventional SNSPDs. Similarly,} sub-nanosecond recovery times can be obtained with extremely short nanowires coupled to integrated waveguides\cite{Vetter2016}, but such detectors have so far not shown high system detection efficiency when coupled to an optical fiber. Arrays of detectors can be used to keep the efficiency up to high detection rates\cite{Rosenberg2013,Zhang2019}\cite{Huang2018}. But this comes at the cost of using multiple coaxial readout lines, which increases the cooling power needed to operate the system, and also demands a proportionally scaled and complex electronics readout circuitry with multiple discriminators and time-tagging units. This limits the scalability of such systems. Row-columns array designs can reduce the number of coaxial lines needed, but still require at least one readout line per row and per column\cite{Allman2015, Wollman2019}, as many amplification channels, and additional processing to manage the timing-accurate readout of those different channel. Moreover, large inductors are needed to prevent crosstalk between pixels, which slows down their individual detection rate. RF-biased SNSPDs\cite{Doerner2017} allow for efficient multiplexing of detectors onto a single coaxial line, however the detection efficiency obtained with and RF bias seems to suffer from the oscillating biasing current, and the complexity of RF signal generation is significantly higher than for conventional DC-biased SNSPDs. Increasing the detection rate of SNSPDs without increasing i) the number of coaxial lines, ii) the constraints on the size and cooling power of the cryostat, and iii) the complexity and costs of the operating and readout electronics, is therefore of great practical advantage. This would contribute to spreading the use of SNSPDs further. 

\begin{figure}
\includegraphics[width=\columnwidth]{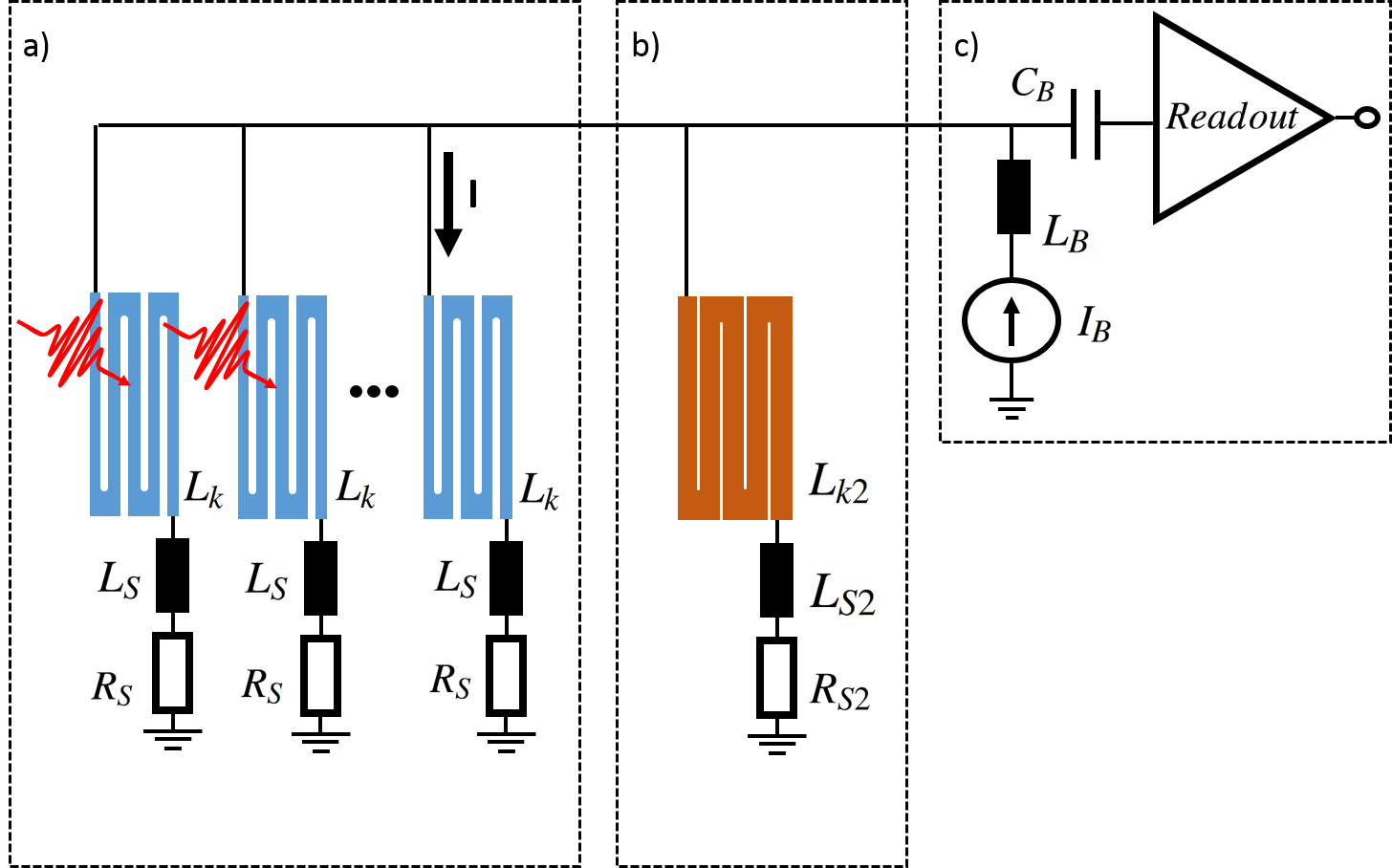}
\caption{\label{fig:schem}Parallel SNSPD design. a) Schematic of a basic parallel SNSPD design\cite{Korneev2007}, which consists of a limited number of photosensitive nanowires with a specific inductance $L_k$ connected in parallel. An additional series inductor $L_s$ can be added to choose the overall inductance of each section which has an impact on the output signal amplitude. Series resistors $R_s$ ensure that the biasing current is evenly split among every nanowire. b) Additional nanowires (only one is shown here)  with  low inductance $L_{k2}$ are added in order to decrease the electronic crosstalk between the nanowires during detection events. The values of $L_{S2}$ and $R_2$ can be sized to minimize the crosstalk while keeping output signal of sufficient amplitudes. c) A bias tee is used to bias the detector and amplify the output signal with the same coaxial line.}
\end{figure}

A potential solution is to use a parallel SNSPDs design, which consists of several nanowires connected in parallel\cite{Korneev2007}, as represented on Fig.~\ref{fig:schem}a. Series resistors ensure that the biasing current is split evenly between the different nanowires. Splitting the whole detector in several nanowires can effectively reduce their individual lengths, hence lowering their respective kinetic inductance, which leads to a shorter recovery time for each nanowire. Moreover only part of the detector is disabled after a detection event which leaves the remaining nanowires available to detect another photon at their full detection efficiency. Similar designs have also been used to demonstrate photon-number-resolving detection \cite{Divochiy2008}, fast recovery time and ultrashort output pulses\cite{Korneev2007, Tarkhov2008}. However, to the best of our knowledge, their potential for achieving larger detection rates has not been fully explored. Indeed, the above mentioned works on ultrafast SNSPDs report measurements at a counting rate of only 100~kHz\cite{Korneev2007, Tarkhov2008} and an efficiency of 0.01\%\cite{Tarkhov2008}. As we observed and report on herein, the problem seems to be that at high detection rate, electronic crosstalk between the different nanowires cumulates, and this leads to a cascading effect between the different nanowires. Ultimately, this can lead to latching, i.e.~all nanowires can end up in a steady resistive state and the whole detector is effectively disabled.

In this work, we first tested parallel designs comparable to those presented in the literature\cite{Korneev2007, Tarkhov2008} and we found that these designs latch at moderate detection rates. We then demonstrate a fiber-coupled parallel SNSPD design that overcomes this limitation. Cascading effect between the nanowires is mitigated by adding carefully designed superconducting nanowires to the structure. Using the same electronic readout as with conventional SNSPDs and a single coaxial line, we demonstrate detection rates over 200~MHz without any latching, and a fibre-coupled SDE as high as 77\%, and more than 50\% average SDE per photon at 50~MHz detection rate under continuous wave illumination.

\begin{figure}
\includegraphics[width=\columnwidth]{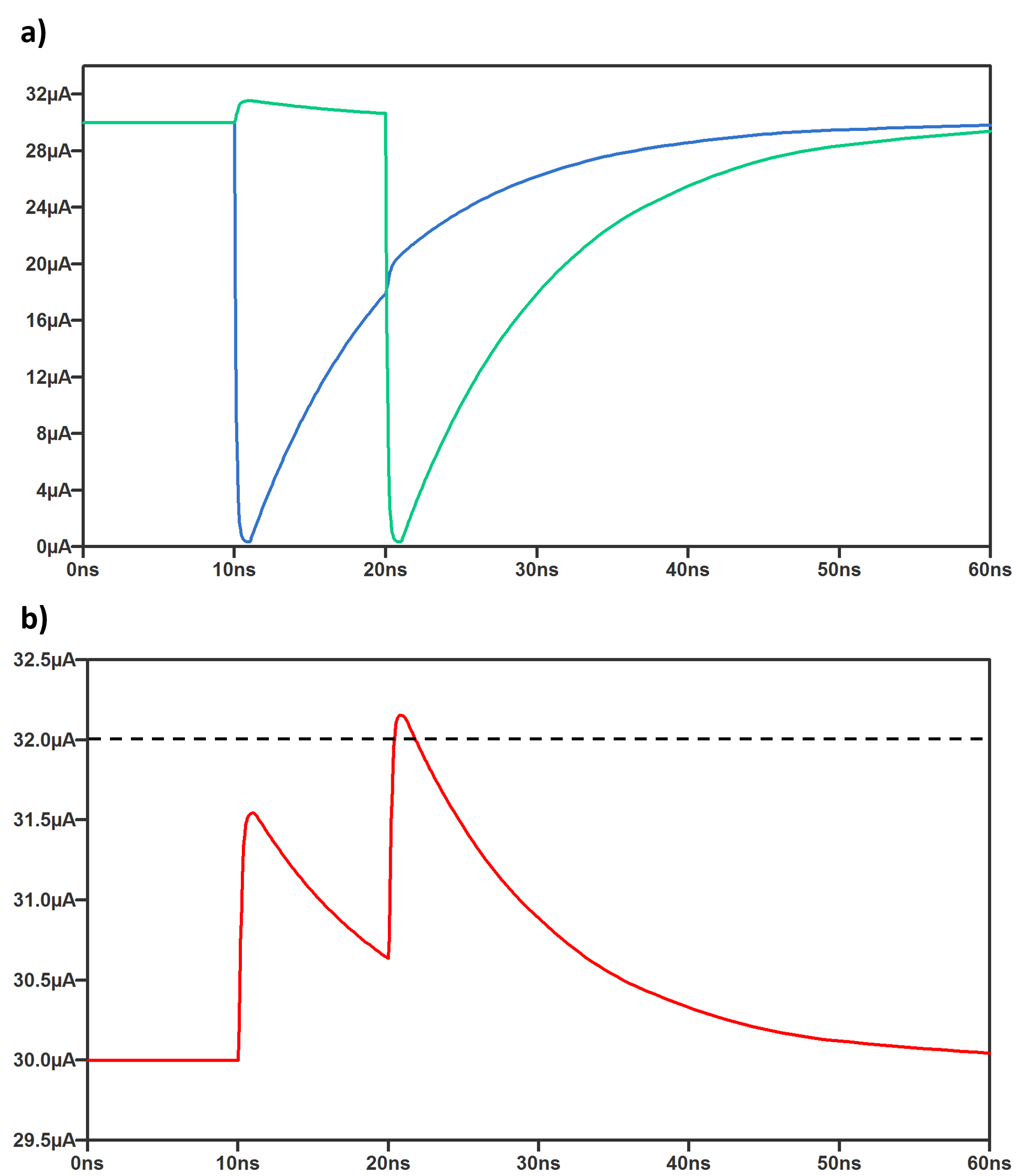}
\caption{\label{fig:Simu} Simulation of the bias current behavior in case of two consecutive detections with a 20-nanowire parallel detector, with hypothetical value of $I_b = 30$~\textmu A per nanowire, and $I_c = 32$~\textmu A, which illustrates the effect of the cumulated electronic crosstalk. a) Bias current in the two nanowires absorbing a photon at 10~ns interval. b) Bias current in the remaining 18 nanowires. For an hypothetical critical current $I_c$ of 32~\textmu A (dashed line), this two photon event causes the remaining nanowires to switch to their resistive state due to the electronic crosstalk.}
\end{figure}

To design a parallel SNSPD for operation at high detection rates, we must consider the electronic crosstalk between the different nanowires : after a photon absorption in one nanowire, the electronic current in that nanowire is partly redirected to every other nanowires as well as to the readout circuit. In each nanowire, this redirected current adds up with the already present bias current\cite{Korneev2007}. At high detection rates, when multiple absorptions occur in different nanowires at time intervals shorter than the recovery time of each nanowire, this effect stacks up and it can eventually bring the current in the remaining nanowires above their local critical current $I_c$ (we assume $I_c$ is the same for each nanowire). This cascading effect can then lead to a latched state. For a given bias current per nanowire $I_b$, only a certain number of photons can then be detected at short time intervals to avoid latching. Simulation results help illustrate this effect; see Fig.~\ref{fig:Simu}. This can be avoided, at least party, by reducing $I_b$ so that a higher number of photons can be detected simultaneously without causing latching for the same time interval. Therefore for a given detection rate, the probability of cascading effect can be made negligibly small through a sufficient reduction of $I_b$, and this should allow for a higher detection rate without latching. However, such a reduction of $I_b$ inevitably decreases the detection efficiency of each nanowire. This means that a conventional parallel SNSPDs cannot necessarily operate simultaneously with their highest efficiency and highest detection rate without latching. Adding more nanowires in parallel will decrease the crosstalk for each detection, as the current is split into more paths, which mitigates the effect as well and allows for more consecutive detection. Still, if a large number of nanowires absorb photons, the remaining ones will switch to the resistive state because of the cumulated crosstalk. There is no theoretical limit to the number of nanowires used, but the amplitude of the output voltage pulse will decrease as well\cite{Korneev2007}. In consequence the number of nanowire is ultimately limited by the amplification capabilities and jitter consideration.

The solution we propose here ensures that, in any nanowire, the sum of $I_b$ and the redirected current created by detections in parallel nanowires will never exceed the critical current $I_c$, even when using a bias current $I_b$ large enough to maintain the saturated efficiency of the detector. To achieve this, additional nanowires are added in parallel to the structure as shown in Fig.~\ref{fig:schem}b. They are positioned outside the optical fiber core such that they are unexposed to light. Part of the redirected current after a detection is therefore split into these unexposed nanowires, which effectively reduces the additional current seen by the exposed nanowires. By designing them with larger widths than the exposed nanowires, it can be ensured that even if every exposed nanowire detects a photon, this additional structure will carry the excess current without reaching its critical current value. Thanks to the reduced crosstalk seen by the exposed nanowires, our design effectively allows for high $I_b$ that can get close to~$I_c$. However this comes at a cost: lowering the crosstalk also lowers the current flowing into the readout, which inevitably decreases the output signal amplitude\cite{Korneev2007}. Hence, the number of parallel nanowires and their minimum kinetic inductance is limited by the performance of the amplification electronics.

In these designs, we also need to consider that the heat generated by the detection mechanism can trigger other nanowires located in the vicinity of the absorption. This thermal crosstalk can be avoided with proper spacing of the different parallel nanowires.We measured the thermal crosstalk between two conventional SNSPDs and observed no thermal crosstalk between nanowires separated by a 800~nm gap. Thus we implemented this distance in our designs. As the fraction of the area dedicated to the gaps increases with the number of exposed nanowires, designs with a larger number of exposed nanowires have a lower fill factor, which in turn decreases the detection efficiency. There is therefore a trade-off to consider.


\input{table}

\begin{figure}
\includegraphics[width=\columnwidth]{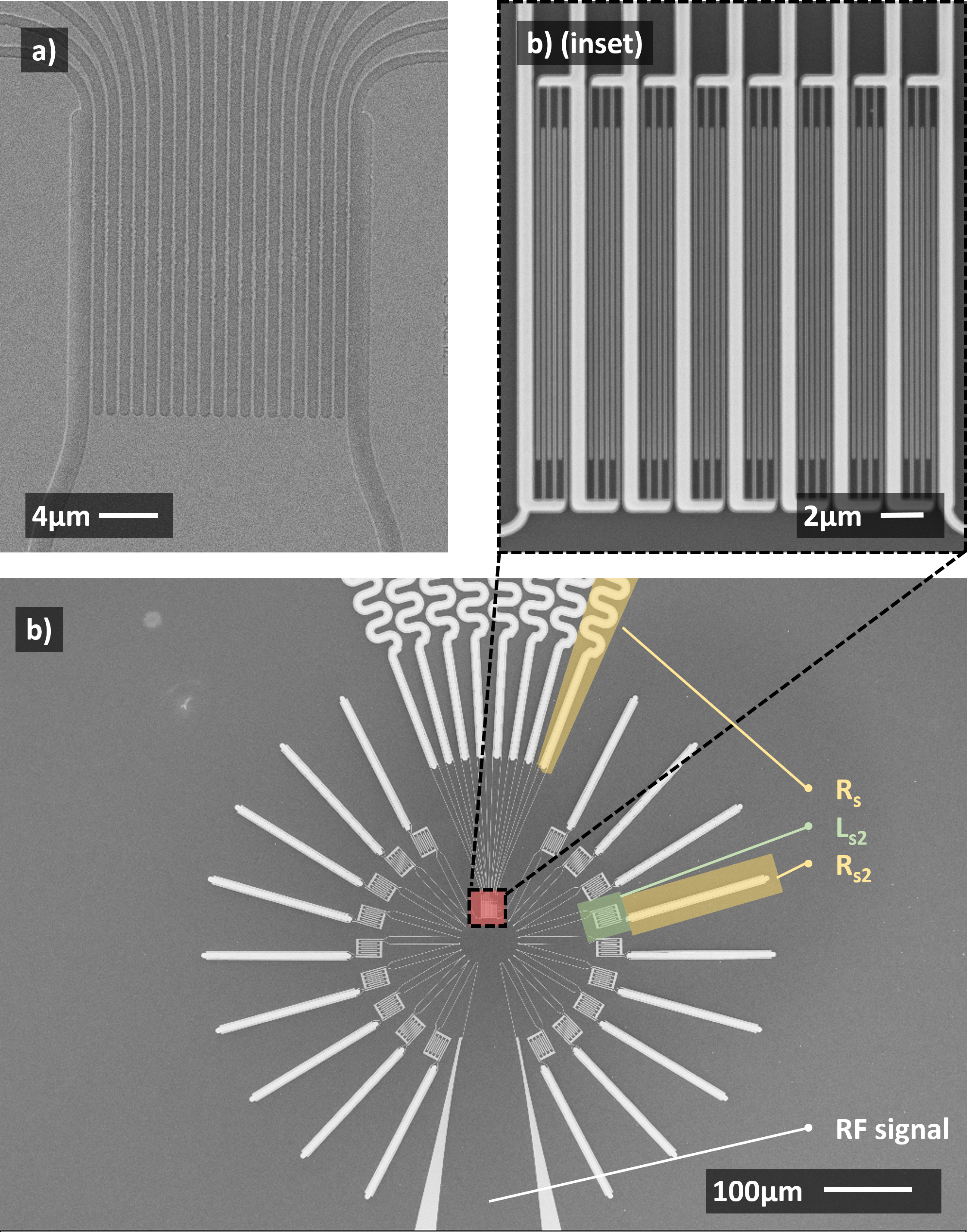}
\caption{\label{fig:SEM}SEM pictures of the devices. a) Zoom on the exposed area of the first solution tested, made of 20 parallel non-meandered nanowires. This design is also similar to device~A, which is made of 40 parallel non-meandered nanowires. In this design, every nanowire has the same width. The superconducting material appears in lighter gray on this SEM picture} b) Overview of a detector similar to device~B, with 8 exposed nanowires and 18 unexposed additional wide wires. The 8 exposed nanowires can be seen in the center of the image (red square) and on the inset. The 18 additional nanowires are positioned on a circle around the center, the green square on the picture highlights one of them. Large gold resistive lines (yellow rectangles) are connected in series with the exposed and unexposed nanowires. The inset shows a zoom on the exposed area of this design, the eight nanowires are shaped in a meander and connected in parallel. The superconducting material appears in darker gray on those two SEM pictures.
\end{figure}

The parallel SNSPDs are patterned using electron beam lithography on 6~nm MoSi film. The bulk superconducting transition of our sputtered MoSi has been measured at 8.3~K. Series resistors and electrodes are created by lift-off of a 10~nm Ti and 90~nm Au evaporated double layer, with photo-lithography techniques used to pattern resistive lines of $5$~\textmu m width and various lengths. Fig.~\ref{fig:SEM} shows a scanning electron microscope (SEM) picture of different parts of the structure. Gold resistors of different values can be seen on the outside, while the photosensitive area (16~\textmu m$\times$16~\textmu m) is found at the center of the image. All devices tested have the same detection area, which correspond to the specification of the optical fiber used. An optical cavity is built around the MoSi layer consisting of a silver mirror and a SiO\textsubscript{2} spacer integrated on the bottom, and a single SiO\textsubscript{2} capping layer on top\cite{Anant2008}. The devices are then separated from the wafer using deep reactive ion etching (DRIE), and packaged using a self-alignment technique\cite{Miller2011}. The packaged SNSPD is cooled down to 0.82~K using a sorption-equipped closed-cycle cryocooler. The detectors are fiber-coupled with telecom single mode fibers. Amplification of the output pulses is done through a first amplification stage cooled at 40~K, and a second amplification circuit at room temperature.

We report on several devices with different geometries to demonstrate the following two points: i) the number of exposed nanowire has an impact on how the efficiency is affected by the detection rate, and ii) additional nanowires with appropriate widths are necessary to prevent latching when the bias current is set close to the critical current and when the detection rates are high. All measurements are performed with continuous light at 1550~nm, at a temperature of 0.82~K.

First, we tested a device made of 20 straight parallel nanowires (Fig. \ref{fig:SEM}a). This design is comparable to previously reported parallel SNSPDs\cite{Korneev2007, Tarkhov2008} and corresponds to the electronic schematics shown in Fig. \ref{fig:schem}a, with $R_s = 5$~Ohms, $L_k \approx 10$~nH and $L_s \approx 100$~nH. This design exhibited clear cascading effect and latched when the overall detection rate was getting close to 10~MHz. This happened even when the bias current $I_b$ was set to the smallest value needed to activate detections, which was $I_b \approx 0.72I_c$. This shows that minimizing the effect of the electrical crosstalk requires a larger number of parallel nanowires. To minimize the electronic crosstalk, we tested a similar design (referred as ``device~A'' later in the text), made of 40~straight identical parallel nanowires, As the optical mode diameter was smaller than the whole structure, only 18 nanowires are exposed to light. However the 40 nanowires all contribute to the reduction of the electronic crosstalk and all of them can suffer from the cumulated electronic crosstalk (Fig \ref{fig:schem}a). with $L_k \approx 10$~nH, $R_s = 5$~Ohms and $L_s \approx 30$~nH. With a current set sufficiently below $I_c$, detection rates as high as 110~MHz were obtained. However, this device still latches at higher detection rates. This latching occurs when several of the exposed nanowires absorb photons at the same time; this re-directs the current partly in the other unexposed nanowires that then reach their critical current values, and this cascading effect leads to a latched state. To prevent this, we here propose and demonstrate a design with only 6~exposed nanowires and 8~wider unexposed nanowires, as shown on Fig.~\ref{fig:schem}b, with values $L_k \approx 188$~nH, $R_s = 20$~ohms, $L_{s2} \approx 23.5$~nH and $R_{s2} = 2.5$~ohms. This design is referred to as "device B" in the text (see Fig.~\ref{fig:SEM}b). The 8~unexposed nanowires have a larger width to ensure that they can carry the diverted current and remain superconducting even in the event of simultaneous detections in all of the 6~exposed nanowires. A summary of the different geometries can be found on table~\ref{tab:devices}

\begin{figure}
\includegraphics[width=\columnwidth]{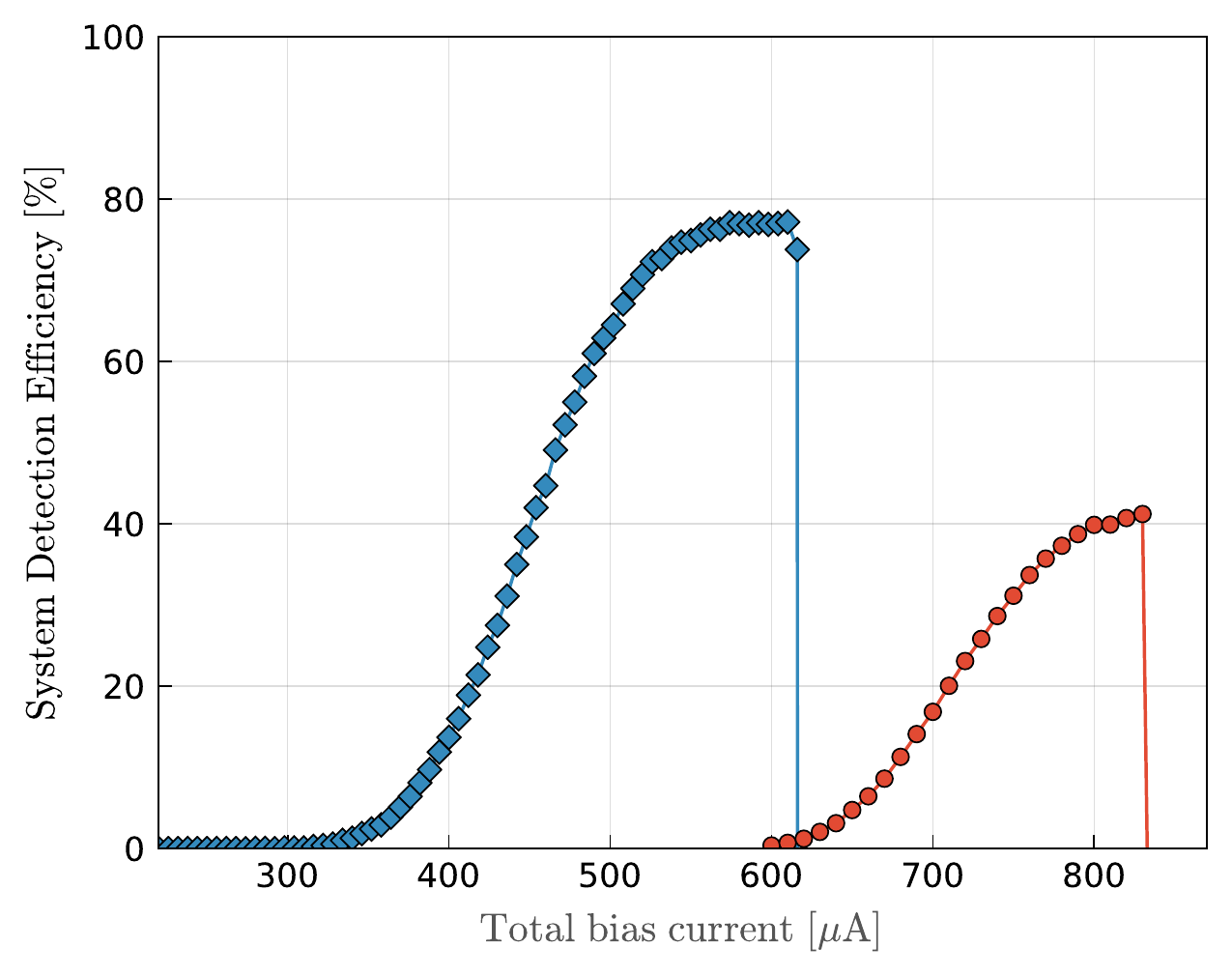}
\caption{\label{fig:Eff} System detection efficiency of device~A (red circles) and device~B (blue squares). The total biasing current used depends on the design due to the different number of nanowires. The efficiency is improved by an integrated optical cavity. A DCR of a few kHz was observed due to blackbody radiation with both designs. This value can be reduced to less than 100~Hz using simple fiber loops to filter unwanted wavelengths, without impacting the overall efficiency\cite{Smirnov2015}.}
\end{figure}

The system detection efficiency of the two different parallel design devices has been characterized using a calibrated powermeter and three variable attenuators\cite{Caloz2018}. The polarization of the incoming light is oriented for maximum detection efficiency. Fig.~\ref{fig:Eff} shows the SDE \textit{vs.}~bias current of both devices at a detection rate of $\sim100$~kHz. An efficiency plateau is reached with a maximum efficiency of 41\% and 77\% respectively. High efficiency is obtained with the second design thanks to a relatively higher fill factor. Indeed, this design consists of nanowires in a meander shape with a fill factor of 60\% and five large gaps of 800~nm between each nanowire, leading to an average fill factor of 42\%. In comparison, device~A is made of 18 exposed wires, all of which are separated by 800~nm gaps. The fill factor of device~A is therefore only 16\% which reduces the maximum efficiency. State of the art SNSPDs can reach efficiencies above 93\%\cite{Marsili2013}. Similar values might be approached by optimizing the gaps of device~B and working on the optical cavity of the device.

\begin{figure}
\includegraphics[width=\columnwidth]{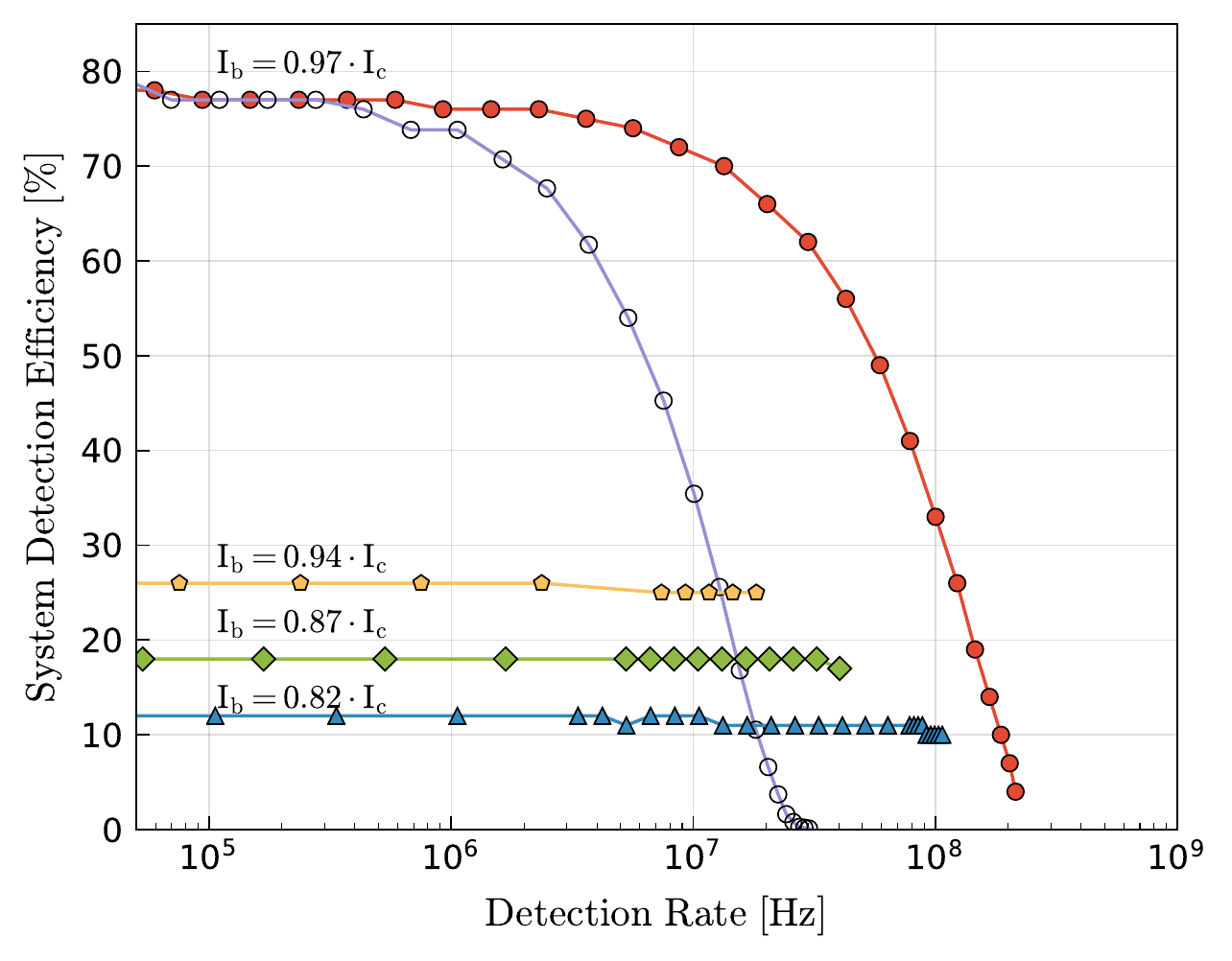}
\caption{\label{fig:EffDrop} System detection efficiency versus detection rate. Red solid circles : Device~B with $I_b = 0.97 \cdot I_c$. Device~B exhibits an SDE at low detection rate almost as high as single meander designs. The device never latches even above 200 MHz. Purple hollow circles : A conventional SNSPD exhibiting loss of efficiency with a detection rate an order of magnitude lower than device B. SDE at low detection rate has been normalized with respect to Device B for better comparison. The blue triangles, green squares, and yellow pentagons curves show the behavior of device~A for different bias currents. Device~A maintains its efficiency at higher detection rates than device~B, but latches when operated at high rates. Decreasing $I_b$ allows for operation at higher detection rates but lowers the efficiency. The measurements have been performed without optimizing the polarization of the incoming light. Hence the efficiencies measured on device~A at low detection rate are slightly lower than the optimal ones expected according to Fig.~\ref{fig:Eff}.}
\end{figure}

As the detection rate increases, the probability of having consecutive photons absorbed in the same exposed nanowire before it fully recovered increases. As a result, the average efficiency per photon does decrease with the rate of incident photons. To characterize this effect, the rate of incident photons illuminating the detector is progressively increased for a given bias current. The detection rate is monitored for each different illumination rate and from this the average SDE is calculated. Fig.~\ref{fig:EffDrop} shows the SDE \textit{vs.}~detection rate of devices~A and B. The operation of device~A at high rates is not possible with a bias current close to $I_c$ because the device is sensitive to the cascading effect, as explained above and shown on~Fig~\ref{fig:EffDrop}. Therefore the bias current $I_b$ needs to be decreased well below $I_c$, which reduces the detector efficiency, namely to 12\% for $I_b = 0.82 \cdot I_c$ as shown on~Fig~\ref{fig:EffDrop}. In this case the detector can operate at detection rates up to 100~MHz before latching. As pointed out already, the large number of exposed nanowires of device~A reduces the probability of fast consecutive absorptions in the same nanowire, and this minimizes impact of the detection rate on the SDE, further than what is observed for device~B. This design exhibits a 10\% efficiency drop at detection rates as high as 88~MHz. An efficiency drop of 10\% of the nominal SDE is observed at $\sim20$~MHz for device~B (see Fig.~\ref{fig:EffDrop}). With this detector, an efficiency above 50\% is obtained at a detection rate of 50~MHz. As desired, the wide unexposed parallel wires protect the device from latching and the detector can still be operated with rates above 200~MHz, with an average efficiency of 6.6\% at 200~MHz. This results confirms the importance of designing wider unexposed parallel wires to protect the detector from cascading and latching, as well as finding a trade-off concerning the number of exposed nanowires to limit the current cross-talk while minimizing the impact of high detection rates on the SDE. We also measured the recovery time of the device using the measurement scheme proposed in our previous work\cite{Autebert2020}. We estimate that each nanowire of device B recovers 90\% of its full efficiency in \textasciitilde 25~ns.

\begin{figure}
\includegraphics[width=\columnwidth]{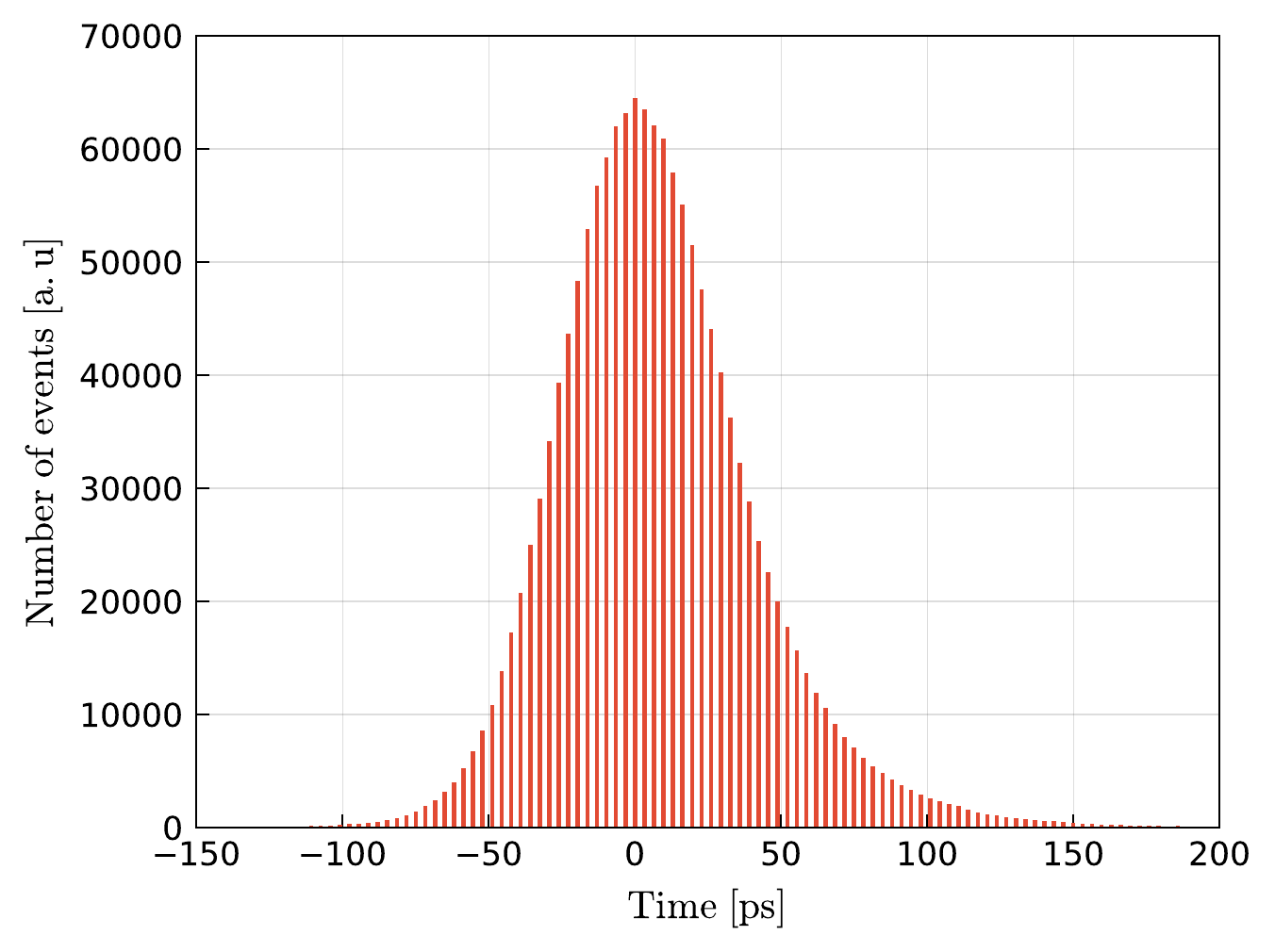}
\caption{\label{fig:jitter} Timing distribution of the photon detection event. Full-width at half maximum is 66.4~ps}
\end{figure}

Another important property of SNSPDs is their timing jitter. We measured a jitter of 66.4~ps full width at half maximum (FWHM) with a detector whose design is similar to device B (Fig.\ref{fig:jitter}). This is higher than what is usually obtained with MoSi single meanders SNSPDs\cite{Caloz2018} and state of the art SNSPDs can exhibit jitter values lower than 15~ns\cite{Zadeh2017, Zadeh2020}. This relatively high jitter could be a consequence of the lower output signal of our devices compared to single meander SNSPDs. Indeed the signal over noise ratio is an important contribution of the total jitter of the system \cite{You2013, Caloz2019}. However, it is interesting to note that the geometric contribution to the jitter per nanowire is expected to be smaller here due to the shorter nanowires required in our designs\cite{Calandri2016, Korzh2018}, which perhaps contribute positively to our jitter value. This measurement was performed at an approximate detection rate of 2~MHz. As with conventional SNSPDs, we expect that the jitter will deteriorate at higher detection rates. Indeed, in this regime consecutive detections in the same nanowire can occur with times intervals that are smaller than the full recovery time of the nanowire, leading to an overall increase of the time jitter.


In conclusion, we fabricated a parallel SNSPD design protected from cascading effects at high counting rates. Comparison with a parallel SNSPDs made of 20 nanowires confirms the necessity of mitigating the electronic crosstalk to operate the device with optimal performance. By integrating in our design several large unexposed nanowires, we created the possibility for current redistribution amongst the detectors, demonstrating latch-free operation up to 200~MHz using a single coaxial line. We also verified the impact of a high number of parallel nanowires to maintain efficiency at even higher counting rates. This work paves the way to push the detection rate of parallel SNPSDs towards the GHz regime, while maintaining high SDE and providing a scalable approach for larger designs. Further improvements in the performances are also foreseen by combining the characteristics of the two designs presented in this work and transitioning to a superconducting material with smaller inductivity.

We acknowledge financial support from the Swiss National Science Foundation (COST project C16.0070) and the SNF NCCR QSIT. We thank Giovanni Resta for useful discussions.

\bibliography{refs.bib}

\end{document}

%% file: table.tex
\begin{table}
\caption{\label{tab:devices}Characteristics of the two devices presented in this work.}

\begin{ruledtabular}
\begin{tabular}{lcc}
 & Device A & Device B\\
\hline
Exposed nanowires & 18 & 6\\
Unexposed nanowires & 22 & 8\\
Width of exposed nanowires & 150 nm & 100 nm\\
Width of unexposed nanowires & 150 nm & 1600 nm\\
Protected from cascade effect & No & Yes\\
\end{tabular}
\end{ruledtabular}
\end{table}